# The cradle of pyramids in satellite images

Amelia Carolina Sparavigna
Dipartimento di Fisica,
Politecnico di Torino, Torino, Italy

We propose the use of image processing to enhance the Google Maps of some archaeological areas of Egypt. In particular we analyse that place which is considered the cradle of pyramids, where it was announced the discovery of a new pyramid by means of an infrared remote sensing.

Saqqara and Dahshur are burial places of the ancient Egypt. Saqqara was the necropolis of Memphis, the ancient capital of the Lower Egypt. This place has many pyramids, including the well-known step pyramid of Djoser, and several mastabas. As told in Wikipedia, 16 Egyptian kings built pyramids there and the high officials added their tombs during the entire pharaonic period [1]. The necropolis remained an important complex for non-royal burials and cult ceremonies till the Roman times. Dahshur is another royal necropolis located in the desert on the west bank of the Nile [2]. The place is well-known for several pyramids, two of which are among the oldest and best preserved in Egypt. Therefore this site can be properly considered as the cradle of Egyptian pyramids [3]. Figure 1 shows the Djoser pyramid and the Great Enclosure at Saqqara. The two images have been obtained from Google Maps after an image processing with two programs, AstroFracTool, based on the calculus of the fractional gradient, and the wavelet filtering of Iris, as discussed in Ref.4. The reader can compare the images with the original Google Maps, using the coordinates given in the figure [5].

Recently the BBC announced the discovery in the area between Saqqara and Dashur, near the river Nile, of a new pyramid buried in the sand [6]. The pyramid has been observed by means of the infrared remote sensing. According to the images in Ref.7, it is located near the pyramid of Khendjer, discovered by Gustave Jequier in 1929, built as the tomb of king Khendjer, who ruled Egypt during the 13th Dynasty [8]. The pyramid currently lies in ruins, in part damaged during the excavations done by Jequier, and it is rising just one meter above the desert sand. The pyramid complex was enclosed by inner and outer walls. The inner wall was made of limestone, the outer wall was made of mud bricks. According to Wikipedia, the pyramid stood at about 37.35 meters high [8].

It is interesting to observe the structure of this pyramid from the space with Google Maps. After a processing by means of AstroFracTool and Gimp [4], we obtained the lower panel in Fig.2. According to Ref.[8], the ruins are rising only a few meters above the grounds; the Google Maps, however, after a suitable processing are displaying all the details of the Khendjer complex.

As BBC announced, Sarah Parcak, of the University of Alabama, used some data from NASA infrared equipped satellites to survey the Egypt. Waiting for a more detailed report on her researches and on the methods the team used, let us observe the images that some Web sites are publishing, in particular that of the Khendjer complex (for the author it is impossible to tell whether the images are the original infrared ones or not). According to [7], it is in this complex that there is one of the discovered pyramids. The site is shown in Fig.3, where the upper panel is displaying as it appears in Google Maps, and the lower panel shows it after processing by means of AstroFracTool. It seems a ghost image having the same features of the complex outlines in Fig.2.

According to the Egypt's Minister of State for Antiquities Affairs, Zahi Hawass, the new technologies are able to locate the remains beneath the sand [9], but it is necessary to identify them with archaeological researches on the spot. Of course, a research on the area will be able to tell the name of the king buried in the site.

As discussed in Ref.10, there are several remote sensing techniques that can be useful in archaeology. For what concerns the Google Maps, let me remark its use in the study of the Merowe Dam and the paleochannels of the Nile where we compared the images from SIR-C/X-SAR imaging radar system, with those from Google [11].

It is my opinion that the image processing of Google Maps can be used for an archaeological survey of Egypt (see for instance, some examples on the satellite images of Amarna, [11]), besides of course, all the satellite methods used for geophysical researches. For what concerns the proposed processing of Google Maps, it is important to note the following fact: it is during the processing activity, when the user is changing parameters and details appear in the picture, that it is easier to recognize them. The information is already in the image: it is only enough to take it out.

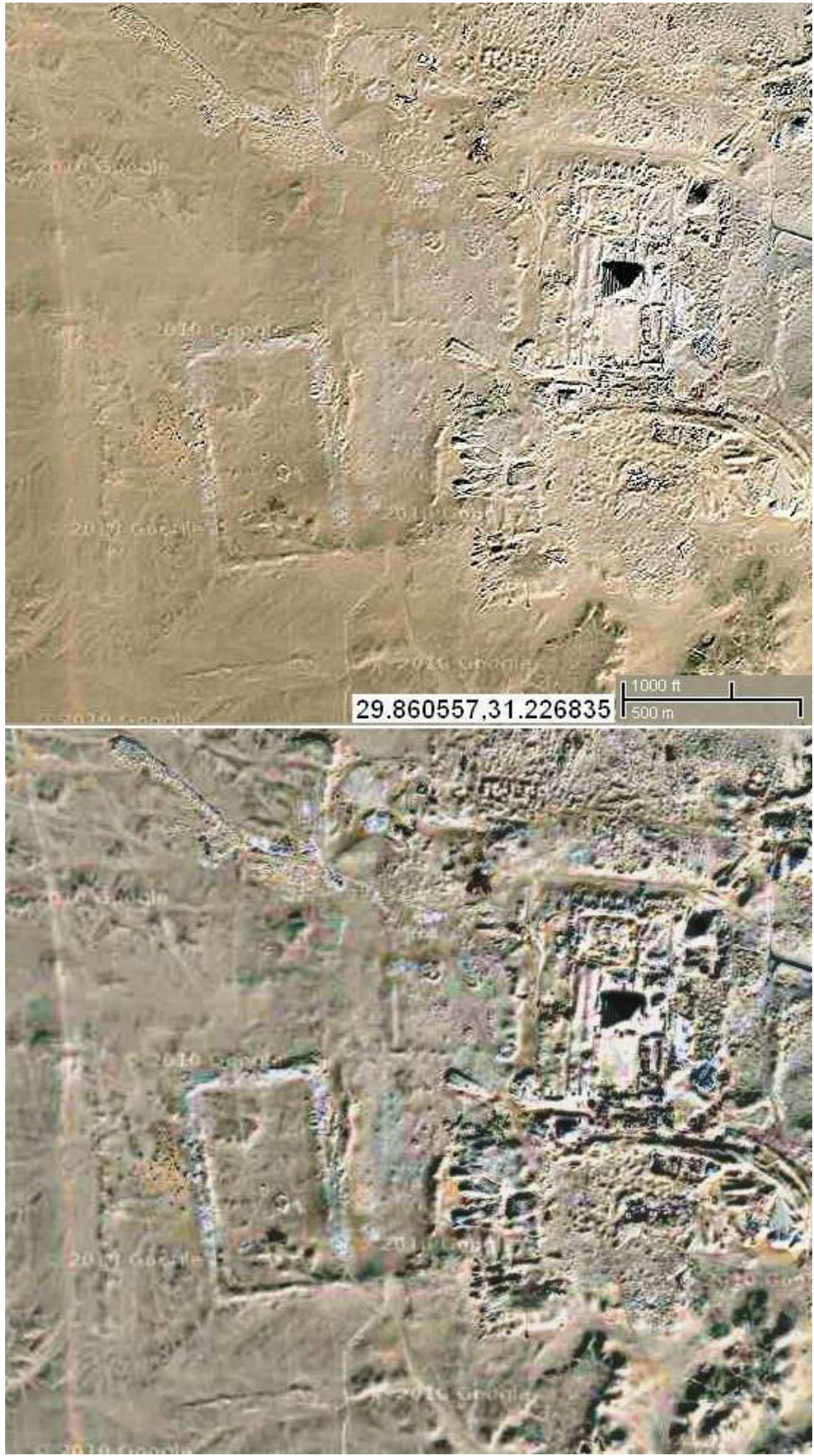

Fig.1: The Djoser pyramid and the Great Enclosure at Saqqara, Egypt. The images have been obtained from Google Maps after a processing by means of AstroFracTool and Gimp, in the upper panel, and Iris wavelets, in the lower panel. For a discussion of the processing see Ref.4. The figure is giving the coordinates for a comparison with the original Google Maps' image [5].

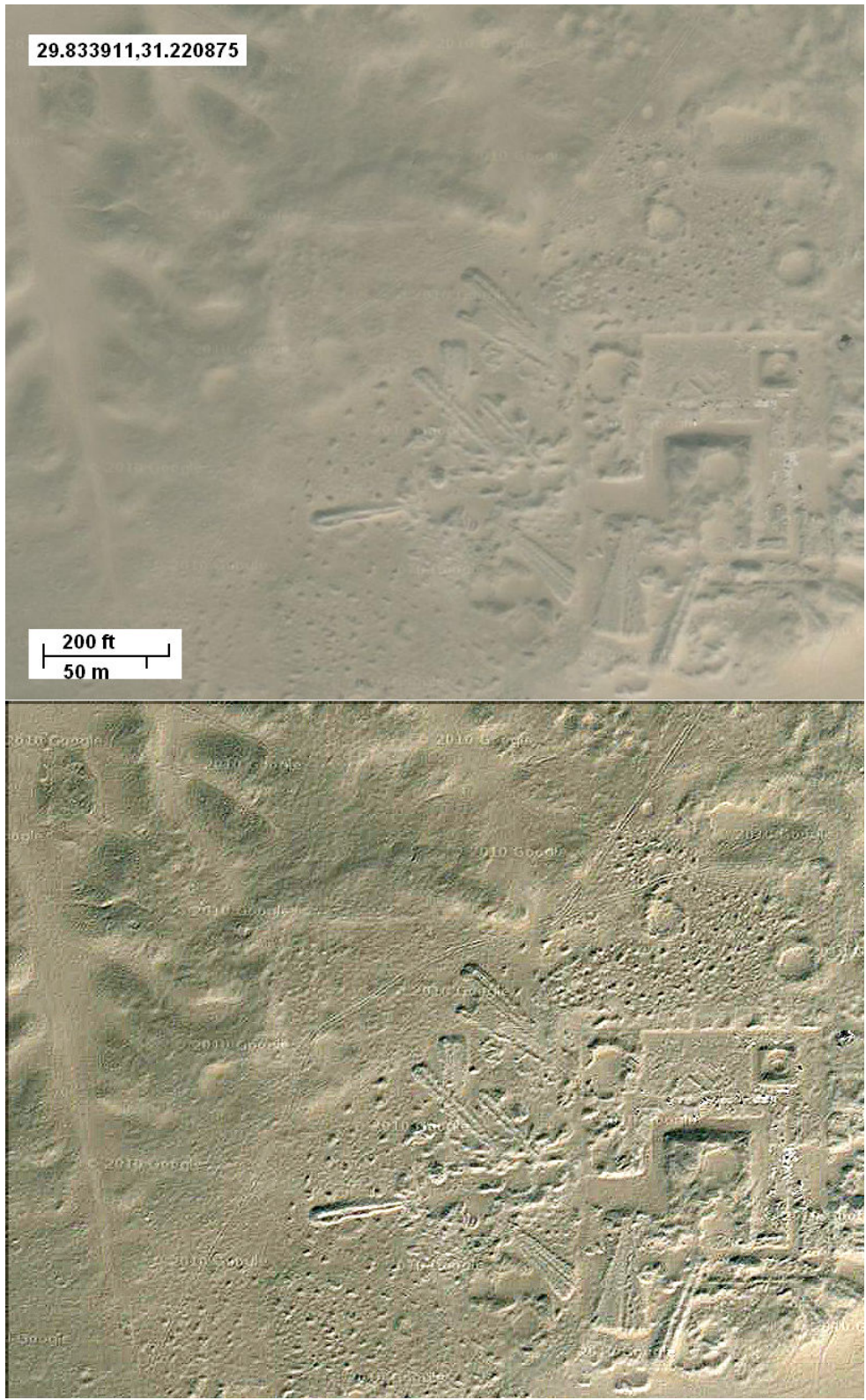

Fig.2 The Khendjer complex as seen from the space, with Google Maps. After a processing by means of AstroFracTool and Gimp [4], we have the lower panel. According to Ref.[8], the ruins are rising only a few meters above the grounds; Google Maps, however, after a suitable processing are displaying all the details.

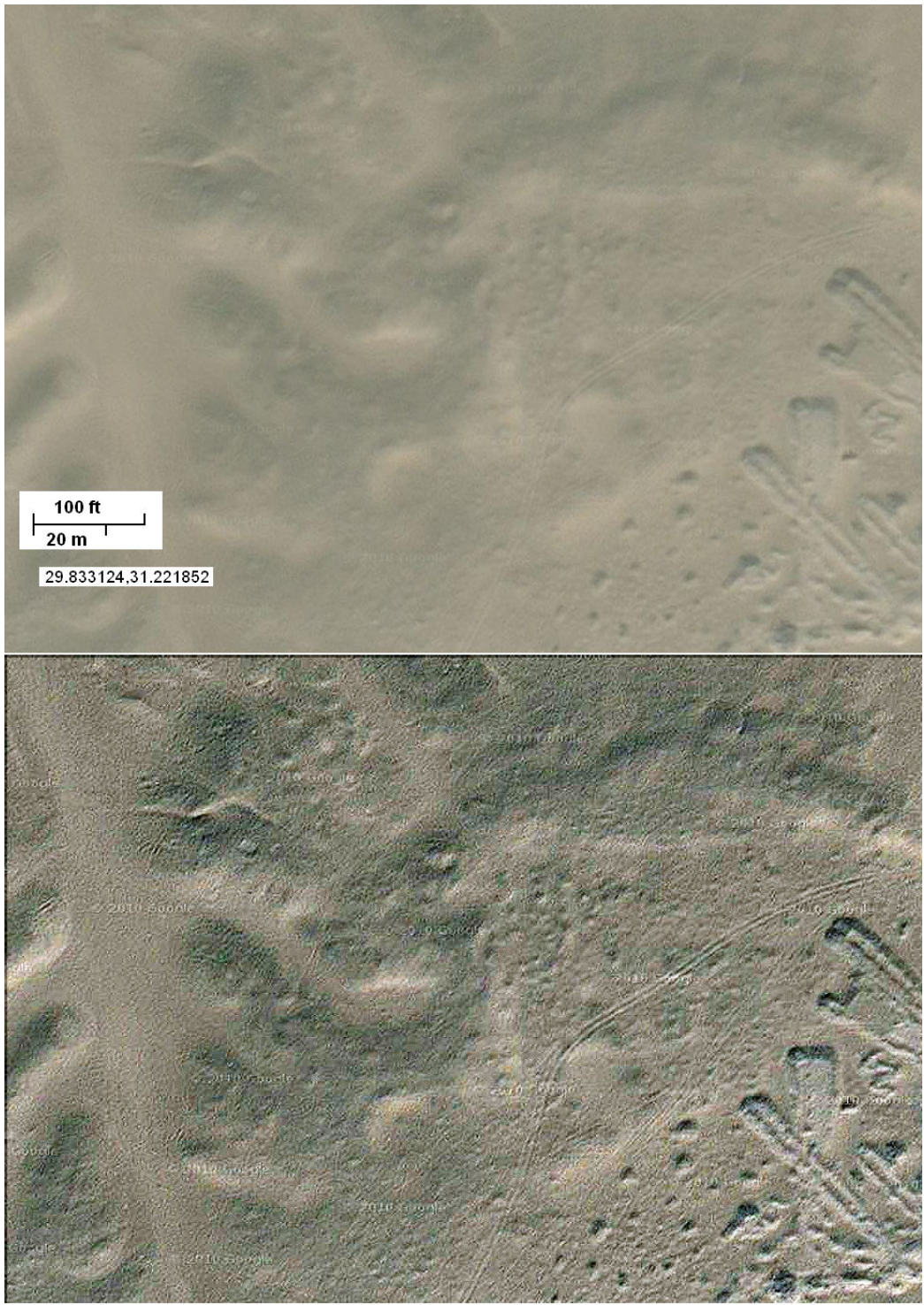

Fig.3. According to [7], it is in the Khendjer complex that there is one of the discovered pyramids. The upper panel is showing as it appears in Google Maps, the lower panel shows it after processing by means of AstroFracTool. It seems a ghost image having the same features of the complex outlines in Fig.2.